\documentclass[12pt]{article}
\usepackage{amsmath}
\usepackage[dvips]{graphics}
\DeclareFontFamily{OT1}{rsfs}{}
\DeclareFontShape{OT1}{rsfs}{m}{n}{ <-7> rsfs5 <7-10> rsfs7 <10->
rsfs10}{} \DeclareMathAlphabet{\mycal}{OT1}{rsfs}{m}{n}
\def\scri{{\mycal I}}%

\begin{document}
\newcommand{\bea}{\begin{eqnarray*}}
\newcommand{\eea}{\end{eqnarray*}}
\newcommand{\bean}{\begin{eqnarray}}
\newcommand{\eean}{\end{eqnarray}}
\newcommand{\eq}[1]{Eq. (\ref{#1})}
\newcommand{\eqs}[1]{Eqs. (\ref{#1})}
\newcommand{\meq}[1]{(\ref{#1})}
\newcommand{\eqn}[1]{(\ref{#1})}

\newcommand{\tri}{\delta}
\newcommand{\grad}{\nabla}
\newcommand{\pa}{\partial}
\newcommand{\pf}[2]{\frac{\pa #1}{\pa #2}}
\newcommand{\bm}{\mathbf}
\newcommand{\de}{\delta}
\newcommand{\pp}{\frac{\partial}{\partial p}}
\newcommand{\pE}{\frac{\partial}{\partial E}}
\newcommand{\ppa}{\frac{\partial}{\partial p'}}
\newcommand{\pEa}{\frac{\partial}{\partial E'}}

\newcommand{\om}{\omega}
\newcommand{\omo}{\omega_0}
\newcommand{\ep}{\epsilon}
\newcommand{\nonu}{\nonumber}
\newcommand{\scrip}{\scri^{+}}
\newcommand{\hp}{{\cal H^+}}

\newcommand{\lxi}{{\cal L}_\xi}
\newcommand{\lt}{{\cal L}_t}
\newcommand{\lchi}{{\cal L}_\chi}
\newcommand{\psig}{{\partial\Sigma}}
\newcommand{\dbar}{\bar\delta}
\newcommand{\La}{\Lambda}
\newcommand{\sh}{S_{\cal H}}

\title{\Large\bf Position space of Doubly Special Relativity}
\author{Sijie Gao${}^1$\footnote{Email: sijie@fisica.ist.utl.pt} and
Xiaoning Wu${}^{2,\ 3}$\footnote{Email:
wuxn@webmail.phy.cnu.edu.tw}\\
1. Centro Multidisciplinar de Astrof\'{\i}sica - CENTRA,\\
Departamento de F\'{\i}sica, Instituto Superior T\'ecnico,\\
Av. Rovisco Pais 1, 1049-001 Lisboa, Portugal \\
2. Department of Physics, National Central university,\\
Chung-li 32054, Taiwan, China\\
3. Max-Planck-Institut f\"ur Gravitationsphysik,\\
Am M\"uhlenberg 1, 14476 Golm, Germany}
\maketitle

\begin{abstract}
We consider three possible approaches to formulating coordinate
transformations on position space associated with non-linear
Lorentz transformations on momentum space. The first approach uses the
definition of velocity and gives the standard Lorentz
transformation. In the second method, we translate the behavior in
momentum space into position space by means of Fourier
transform. Under certain conditions, it also gives the
standard Lorentz transformation on position space.
The third approach investigates the covariance of
the modified Klein-Gordon equation obtained from the dispersion
relation.

\end{abstract}
\section{Introduction}
Doubly Special Relativity (DSR) has been investigated extensively in the
last three years \cite{gio1}\cite{gio2}\cite{gio3}\cite{kow1}\cite{kow2}. When
quantum effect is taken into account, classical relativity is no longer
sufficient to describe spacetimes. For instance, Padmanabhan\cite{TP}
  argued that combining gravity with quantum theory prevents the
  measurement of a single event with an accuracy better that the Plank
  length $l_p$. The Plank length/energy is obviously expected to
play a role in quantum gravity. Special relativity
suggests that if the scale $l_p$ is measured in one inertial reference frame,
it may be different  in another observer's frame. So one faces a
direct conflict when Plank scale is introduced to special relativity.
To solve this paradox, DSR theories modify special relativity with two
observer-independent scales. For example,  Magueijo and Smolin
\cite{prl} proposed a non-linear representation of Lorentz group on
momentum space such that the Planck energy is left invariant.
Specifically, the non-linear representation of the Lorentz group
takes the form
\bean
W=U^{-1}LU, \label{orl}
\eean
where $L$ is the ordinary Lorentz generator acting on momentum
space and $U$ is defined by

\bean
U\circ p_a=\frac{p_a}{1-l_p p_0}. \label{defu}
\eean
Consequently, the boosts in the $x$ direction are given by
%*************************olt olx oly olz
\bean
p_0'=\frac{\gamma(p_0-v p_x)}{1+l_p(\gamma-1)p_0-l_p\gamma v p_x}
\label{olt} \\
p_x'=\frac{\gamma(p_x-v p_0)}{1+l_p(\gamma-1)p_0-l_p\gamma v p_x}
\label{olx} \\
p_y'=\frac{p_y}{1+l_p(\gamma-1)p_0-l_p\gamma v p_x}
\label{oly} \\
p_z'=\frac{p_y}{1+l_p(\gamma-1)p_0-l_p\gamma v p_x}.
\label{olz}
\eean
A general isotropic $U$-map discussed by Magueijo and Smolin \cite{gen} takes
the form
%********************genu
\bean
(E', {\bf p'})= U\circ(E,{\bf p})=(Ef_1(E), {\bf p} f_2(E)), \label{genu}
\eean
Note that a general invariant quantity associated with the group
action \eq{orl} is \cite{gen}
%********************mop
\bean
||p||^2\equiv \eta^{ab}U(p_a)U(p_b) \label{mop}
\eean
If the invariant $||p||^2$ is identified with the square of the mass
$||p||^2=m_0^2$, one obtains the general isotropic dispersion relation
%*****************************disp
\bean
E^2 f_1^2-p^2 f_2^2=m_0^2, \label{disp}
\eean
where $p^2=|{\bf p}|^2 $.
For $U$ given in \eq{defu}, we have
%**********************dis
\bean
\frac{E^2-p^2}{(1-l_p p_0)^2}=m_0^2 . \label{dis}
\eean

To make the Planck energy $E_p$ invariant under the action of the
Lorentz group, $U$ must be singular at $E_p$. Conservation of energy
and momentum and other properties concerning the modified Lorentz
transformation have been explored in \cite{gen} and  \cite{matt}. The
main interest of this paper is to investigate the consequences of such
a transformation on position space. One non-linear Lorentz
transformation on position space analogous to \eq{orl} is
%*******************************vc
\bean
K=(U^*)^{-1}LU^* ,\label{vc}
\eean
where $L$ is the standard Lorentz transformation and $U^*$ acts on
position space. If $U^*$ is taken as \cite{posi}
%*********************************vx
\bean
U^*(x^\mu)=x^\mu\frac{1}{t+R}, \label{vx}
\eean
it leads to the Fock-Lorentz transformation \cite{fock}\cite{man}
%******************************ft fx fy fz
\bean
t'&=&\frac{\gamma(t-vx)}{1-(\gamma -1)t/R+\gamma vx/R} \label{ft} \\
x'&=&\frac{\gamma(x-vt)}{1-(\gamma -1)t/R+\gamma vx/R} \label{fx} \\
y'&=&\frac{y}{1-(\gamma -1)t/R+\gamma vx/R} \label{fy} \\
z'&=&\frac{z}{1-(\gamma -1)t/R+\gamma vx/R} \label{fz}.
\eean
However, this treatment only provides an analogy to the transformation
on momentum space. An open question is what transformations on
position space are compatible with the non-linear transformations on
momentum space above. We shall explore this issue in three distinct
ways. First, we use velocity as a link to connect the momentum
space and position space.  There are different proposals on the
definition of velocity in DSR theories
\cite{taka}\cite{mig}\cite{vprd}\cite{vel}\cite{giov}.
In special relativity, the
group velocity  $v_g=\pf{E}{p}$  is equal to the boost
velocity. However, in DSR theories,   $v_g$ is always  mass-dependent
\cite{mig}. For example, in the  Magueijo-Smolin model above, the group
velocity reads,
%*******************cv
\bean
v_g=\frac{v\gamma}{\sqrt{2l_p m_0\gamma+\gamma^2+l_p^2 m_0^2}} ,
\label{cv}
\eean
where $v$ is the boost velocity. This seems to be odd since the
velocity of a particle depends on its
mass. Kosi\'nski and Ma\'slanka \cite{vprd}demand that the velocity be
a property
of reference frame rather than of a particular object and then, the
velocity is identified with the boost velocity $v$, as in the case of
special relativity. From this assumption and the fact that the group
structure remains the same as in Einstein's theory, the authors of
\cite{vprd} derived the ordinary relativistic velocity law for a
general DSR theory. Daszkiewicz, {\em et al.} \cite{vel} investigated the
velocity of particles in
DSR, defining velocity as the Poisson bracket of position with
appropriate Hamiltonian:
%*************************fvo fvi
\bean
u_0&\equiv& \dot x_0 =[x_0, {\cal H}] \label{fvo}\\
u_i&\equiv& \dot x_i=[x_i, {\cal H}] \label{fvi}
\label{fv}
\eean
where $(u_o, u_i)$ is the four velocity of particle. They also found that
the four velocities transform as standard Lorentz four vectors and
the boost parameter $\xi$ is related to velocity in exactly the same
way as in the Special Relativity, i.e. $v=\tanh\xi$.  Based on these
works , we also identify the boost velocity $v$ as the true
velocity in DSR theories and derive
the classical relativistic addition law. Most
importantly, from the original definition $v=dx/dt$, we show that the standard
Lorentz transformation on position space is inferred by the velocity
addition law, despite the fact that the momentum transformation is non-linear.

In section \ref{FT}, we follow Sch\"utzhold and Unruh's prescription
 to induce a transformation on
position space. Fourier transform plays an important role in this
 method. By requiring some suitable conditions,  we obtain the
 standard Lorentz transformation again.

The third method to
fix the transformation law in position space is making use of the
covariance requirement on the modified Klein-Gordon equation (see
\eq{rt}). We show that the only linear transformations that keep
\eq{rt} invariant are pure rotations, i.e., boosts are ruled out. This
may indicate that the modified K-G equation is
associated with  a preferred inertial frame. An alternative resolution
is that the coordinate transformation inferred by the modified K-G
equation is not linear.  Kimberly {\em et. al} \cite{posi} also
required that $p_a dx^a$
remain invariant and derived an energy-dependent boost in position
space (See \eq{ilt} and \eq{ilx}). But this result contradicts
the one we obtain from the covariance of the field
equation. Therefore, the condition that $p_a dx^a$ remains invariant
may not be valid in DSR theories.

For simplicity, we shall work in $1+1$ dimensions. There should be no
difficulties to generalize our results to four dimensions.

\section{Transformations on position space}

\subsection{Velocity addition and coordinate transformation}
As discussed in the introduction, we treat the boost velocity $v$ as
the velocity measured by an inertial reference frame.
We first show that the non-linear transformation \meq{orl} indicates the
ordinary relativistic velocity addition law. Since each Lorentz
transformation $L$ is determined by the relative velocity $v$, we
rewrite \eq{orl} as
%*************************************ru
\bean
W(v)=U^{-1}L(v)U \label{rul},
\eean
We now show that  the standard Lorentz transformation
%****************gumo gumx
\bean
p_0'&=&\gamma(p_0-v p_x)
\label{gumo} \\
p_x'&=&\gamma(p_x-v p_0).
\label{gumx}
\eean
leads to the relativistic addition law for velocities. Let $A$, $B$ and
$C$ be three inertial observers with relative velocities $v_{AB}$,
$v_{BC}$ and $v_{AC}$. Then we have
%***********************************lv
\bean
L(v_{AB})L(v_{Bc})=L(v_{AC}). \label{lv}
\eean
Substituting  \eqs{gumo} and \meq{gumx} into \eq{lv}, we obtain the
relativistic velocity addition law
%*****************************vt
\bean
v_{AC}=\frac{v_{AB}+v_{BC}}{1+v_{AB} v_{BC}}. \label{vt}
\eean
Since a non-linear transformation  is a representation of the
Lorentz group, the  velocity relation \meq{vt} must hold for all
transformations like \eq{rul}.
Next, we shall show that \eq{vt} implies the standard Lorentz
transformation on
coordinate space. Suppose that the coordinate transformation between
two inertial frames is
%***************************gcx gct
\bean
x'&=& g(t,x) \label{gcx} \\
t'&=& f(t,x) \label{gct}
\eean
By differentiation, we have
%***********************dx dt
\bean
dx'&=&\dot g dt+g' dx \label{dx} \\
dt'&=&\dot f dt +f' dx \label{dt}
\eean
Here, $\dot g=\pa g/\pa t$ and $g'=\pa g/\pa x$, etc.
Hence,
%***************************uxp
\bean
u_x'=\frac{dx'}{dt'}=\frac{\dot g +g'u_x}{\dot f +f' u_x}.\label{uxp}
\eean
The relative velocity $v$ between the two frames can be read off by
taking $dx'=0$ in \eq{dx}. Therefore,
%************************rv
\bean
v=-\dot g/g' \label{rv}.
\eean
By comparing \eq{uxp} with \eq{vt}, we obtain the following relations
%*****************thre
\bean
g'=\dot f, \ \ \ \ \dot g/\dot f =-v, \ \ \ \ \ f'/\dot f =-v \label{thre}
\eean
Using these relations, we rewrite \eq{dx} and \eq{dt} as
%***************************rdx rdt
\bean
dx' &=& -v \dot f dt +\dot f dx \label{rdx} \\
dt' &=& \dot f dt -v \dot f dx \label{rdt}.
\eean
The inverse transformation thereby is
%**************************ix it
\bean
dx &=&\dot f^{-1}\frac{v}{1-v^2}dt'+ \dot f^{-1}\frac{1}{1-v^2}dx'\label{ix} \\
dt &=&\dot f^{-1}\frac{1}{1-v^2}dt'+ \dot f^{-1}\frac{v}{1-v^2}dx' \label{it}.
\eean
On the other hand, for symmetry reasons \eqs{ix} and \meq{it} may be
obtained from
\eqs{rdx} and \meq{rdt} by interchanging the primed and the unprimed
variables and replacing $v$ by $-v$, i.e.,
%***************************irdx irdt
\bean
dx &=& v \dot f dt' +\dot f dx' \label{irdx} \\
dt &=& \dot f dt' +v \dot f dx' \label{irdt}.
\eean
By comparing coefficients, we find immediately
%***************************fd
\bean
\dot f =\frac{1}{\sqrt{1-v^2}}=\gamma. \label{fd}
\eean
Therefore, all derivatives of $f$ and $g$ are constants, which
indicates that \eqs{gcx} and \meq{gct} are linear transformations.
By solving \eqs{fd} and \meq{thre}, we obtain the standard Lorentz
transformation on coordinate space.

\subsection{Fourier transform method} \label{FT}
In this section, we shall use Fourier transform to translate the
behavior in momentum space into position space. Sch\"utzhold and
Unruh\cite{unruh} suggest the following way translating a scalar field
$\phi(t,x)$ from one frame to another:
%**************************ueo uet ue
\bean
\phi(t,x)&\rightarrow &\tilde\phi(E,p)={\cal F}\phi \label{ueo}  \\
\tilde\phi(E,p)&\rightarrow &\tilde\phi(E',p')  \label{uet}\\
\phi'(t,x)&=&{\cal F}^\dagger\tilde\phi(E',p') \label{ue},
\eean
where $(E',p')=U\circ(E,{\bf p})$ is the transformation on momentum
space defined by \eq{genu} and ${\cal F}$ is the Fourier
transform. The above prescription enables us to induce a coordinate
transformation $(t,x)\to (U^*t,U^*x)$ determined by
%******************************phic
\bean
\phi'(t,x)=\phi(U^*t,U^*x) \label{phic}.
\eean
It is not difficult to check that the Lorentz transformation can be
recovered this way from the usual Lorentz transformation on momentum
space. However, since we deal with non-linear transformations, it is
not easy to work out a result from a general $\phi(t,x)$. So we shall
replace $\phi(t,x)$ with a single coordinate $t$ or $x$ and by following
Eqs. \meq{ueo}-\eqn{ue}, obtain the coordinate transformation
directly. Hence, \eq{ueo} yields
%**********************************fxt fxx
\begin{eqnarray}
{\cal F}[t]&=&-i\frac{\partial}{\partial E}\de(E)\de(p)\label{fxt} \\
{\cal F}[x]&=&i\frac{\partial}{\partial p}\de(p)\de(E)\label{fxx}.
\end{eqnarray}
According to Eqs. \meq{uet} and \meq{ue}, we change variables from $(E,p)$
to $(E',p')$ and then Fourier transform them back to get new
coordinates on position space. The operations on $t$
and $x$ respectively give rise to the new coordinates:
%*********************************************rtp rxp
\begin{eqnarray}
U^*t&=&\int [-i\pEa\de(E')\de(p')]\exp(ipx-iEt)dpdE \label{rtp} \\
U^*x&=&\int [i\ppa\de(p')\de(E')]\exp(ipx-iEt)dpdE \label{rxp},
\end{eqnarray}
where $(E',p')=U\circ(E,p)$. In order to perform the
integrals, we need to express $(E,p)$ as functions of $(E',p')$.
By definition, we have
%**************************************iu
\begin{eqnarray}
(E,\ p)=U^{-1}\circ(E',p') =\left(h_1(E',p')E',\ h_2(E',p')p'\right)
\label{iu} \eean Hence, \bean det\frac{\partial(E,p)}{\partial
(E',p')}&=&(h_1+\dot h_1E')(h_2+h_2'p')-h_1'\dot h_2E'p' ,
\end{eqnarray}
where $h'$ and $\dot h$ are the derivatives with respect to $p$ and
$E$, respectively. Thus, \eq{rtp} and \eq{rxp} can be rewritten as
%*******************usx
\begin{eqnarray}
U^*x&=&\int [i\ppa\de(p')\de(E')]\exp(ipx-iEt)dpdE\nonumber\\
&=&\int
[i\ppa\de(p')\de(E')]\exp(ih_2p'x-ih_1E't)det\frac{\partial(E,p)}{\partial
  (E',p')}dE'dp'\nonumber \\
&=&-\int
\de(p')\de(E')i\ppa\left[\exp(ih_2p'x-ih_1E't)det\frac{\partial(E,p)}{\partial
    (E',p')}\right]dE'dp'\nonumber \\
&=&h_1(0)h^2_2(0)x-i[h_1'(0)h_2(0)+2h_1(0)h_2'(0)] \label{usx}\\
U^*t&=&\int [-i\pEa\de(E')\de(p')]\exp(ipx-iEt)dpdE\nonumber\\
&=&-\int
\de(E')\de(p')(-i)\pEa\left[\exp(ih_2p'x-ih_1E't)
det\frac{\partial(E,p)}{\partial (E',p')}\right] dE'dp'\nonumber\\
&=&h_1^2(0)h_2(0)t+i[2\dot h_1(0)h_2(0)+h_1(0)\dot h_2(0)] \label{ust}.
\end{eqnarray}
Similarly, we have
\begin{eqnarray}
U^{-1*}x&=&\frac{x}{h_1(0)h_2^2(0)}+\frac{i}{h_1(0)h^2_2(0)}
\left(\frac{h_1'(0)}{h_1(0)}+2\frac{h_2'(0)}{h_2(0)}\right)\\
U^{-1*}t&=&\frac{t}{h^2_1(0)h_2(0)}-\frac{i}{h^2_1(0)h_2(0)}\left(2\frac{\dot
  h_1(0)}{h_1(0)}+\frac{\dot h_2(0)}{h_2(0)}\right)
\end{eqnarray}
(Note : Here we have used the relation $\de(f(x)x)=\frac{1}{f(x)}\de(x)$.)

From the above results, we see that $U^*$ generally is a complex
transformation. However, the imaginary parts will vanish if $h_1$ and
$h_2$ are  even functions of $(E,p)$ and smooth at the origin
$E=p=0$. We shall show later in this
section that this condition is sufficient to preserve the conservation
of electric charge. Therefore, by imposing this condition, we find that
$U^*$ is just a constant dilation, i.e.,
\begin{eqnarray}
U^*x&=&h_1(0)h_2^2(0)x\\
U^*t&=&h^2_1(0)h_2(0)t.
\end{eqnarray}
Substituting it into Eq.(\ref{vc}), we get
\begin{eqnarray}
t'&=&\gamma\left(t-v\frac{h_2(0)}{h_1(0)}x\right)\\
x'&=&\gamma\left(x-v\frac{h_1(0)}{h_2(0)}t\right) .
\end{eqnarray}
As required in \cite{prl}, transformation \meq{orl} should reduce to the
ordinary Lorentz transformation for energy scales much smaller than $E_p$,
which means $\lim_{E\to 0}U=1$, i.e., $h_1(0)=h_2(0)=1$, so we find that the
induced transformation on coordinate space is just the ordinary
Lorentz transformation, which
agrees with our result in section 2.1.

Now we explain why we require $h_1$ and $h_2$ be even functions.
Based on Sch\"utzhold and Unruh's result \cite{unruh}, the
general form of the map $U^*$ is
\begin{eqnarray}
\label{un}
[U^*\phi](x,t)&=&\int B(x,t;\xi,\eta)\phi(\xi,\eta)d\xi d\eta\nonumber\\
B(x,t;\xi,\eta)&=&\int \exp (-ip'\xi+iE'\eta+ipx-iEt)dpdE .
\end{eqnarray}
If $\phi$ is a real scalar field, we wish to see whether $U^*\phi$ is
real too. Based on Eq.(\ref{un}), the complex conjugate of $U^*\phi$
is
\begin{eqnarray}
\label{bar}
\overline{[U^*\phi]}(x,t)&=&\overline{\int
  B(x,t;\xi,\eta)\phi(\xi,\eta)d\xi d\eta}\nonumber\\
&=&\int\int \exp (ip'\xi-iE'\eta-ipx+iEt)\bar\phi(\xi,\eta)dpdEd\xi
  d\eta\nonumber\\
&=&\int\int \exp (if_2p\xi-if_1E\eta-ipx+iEt)\phi(\xi,\eta)dpdEd\xi
  d\eta .
\end{eqnarray}
In general, $\overline{U^*\phi}\ne U^*\phi$, i.e., $U^*\phi$ is
not a real function. If $h_1$ and $h_2$
are even functions, i.e., invariant under the coordinate
transformation $E\to -E$ and $p\to -p$, we have
\begin{eqnarray}
\overline{U^*\phi}&=& \int\int \exp
  (if_2p\xi-if_1E\eta-ipx+iEt)\phi(\xi,\eta)dpdEd\xi d\eta \nonumber\\
&=&\int\int \exp
  [if_2(-E,-p)(-p)\xi-if_1(-E,-p)(-E)\eta\nonumber\\
  &&\qquad -i(-p)x+i(-E)t]\phi(\xi,\eta)dpdEd\xi d\eta \nonumber\\
&=&\int\int \exp
(-if_2p\xi+if_1E\eta+ipx-iEt)\phi(\xi,\eta)dpdEd\xi
  d\eta\nonumber\\
&=&U^*\phi .
\end{eqnarray}
Thus, $U^*\phi$ is real. If $U^*\phi$ is not real, it will cause a serious
problem. It is well known that a complex field contains  electric
charges. If the modified
Lorentz transformation can not guarantee that $U^*\phi$ is real,
different observers will have different views on
whether a particle is charged, which is in contradiction with the
conservation of electric charge. However, we only show that the
even function requirement is a sufficient condition for
$U^*\phi$ being  real. It may not be necessary.

\subsection{Coordinate transformation from field equation}
As outlined in \cite{prl}, the derivatives in a field equation should
transform as momentum. One can construct the modified scalar field
equation by the replacement
%***************************rp
\bean
p_a\rightarrow i\pa_a \label{rp}
\eean
applied to the dispersion relations. Thus, from \eq{dis},
we have the modified Klein-Gordon equation\cite{posi}
%*********************************rt
\bean
\eta^{ab}\frac{\partial_a}{1-il_p\partial_0}\frac{\partial_b}
{1-il_p\partial_0}
\phi(x)=0, \label{rt}
\eean
which has plane wave solutions $\phi=A e^{-ipx}$ with $p_a$ satisfying
the dispersion relation \eq{dis}. A basic requirement for a field
equation is that it must be covariant under coordinate transformation
between inertial frames. Conversely, this requirement can be used to
determine possible coordinate transformations that keep the field
equation \meq{rt} invariant.
The explicit expression of \eq{rt} is obtained by Taylor expansion
%**********************te
\bean
\eta^{ab}\partial_a(1-il_p\partial_0...)\partial_b(1-il_p\partial_0...)
\phi(x)=0 \label{te}.
\eean
It is not difficult to see that in four dimensions,  pure
spatial rotations will make
\eq{te} invariant. Now we show that spatial rotations actually are the only
linear transformations on position space that make \eq{te} invariant.
Without loss of generality, we still consider $1+1$ spacetimes and
\eq{te} becomes
%***************************fe
\bean
-\pa_t^2 \phi(x)+\pa_x^2 \phi(x)-2il_p\pa_t^3\phi(x) +2il_p\pa_x^2\pa_t
\phi(x)+... =0.
\label{fe}
\eean
Consider coordinates $(t',x')$.  From the chain rule, we have
%***************************pt px
\bean
\pa_t&=&\pf{t'}{t}\pa_t'+\pf{x'}{t}\pa_x' \label{pt} \\
\pa_x &=&\pf{t'}{x}\pa_t'+\pf{x'}{x}\pa_x' \label{px}.
\eean
Substituting \eq{pt} and  \eq{px} into \eq{fe}, we get
%***********************************lp
\bean
&&-\pa_t^2 \phi(x)+\pa_x^2 \phi(x)-2 il_p\pa_t^3\phi(x)
+2il_p\pa_x^2\pa_t \phi(x)  +...   \nonumber \\
&=&\left[-(\pf{x'}{t})^2+(\pf{x'}{x})^2+...\right]\phi''+\left[-(\pf{t'}{t})^2+
(\pf{t'}{x})^2+...\right]\ddot\phi+\left[-\pf{x'}{t}\frac{\pa}{\pa x}\left(
  \pf{x'}{t} \right)+...\right]\phi' \nonumber \\
&&+\left[-\pf{x'}{t}\frac{\pa}{\pa
    x}\left(\pf{t'}{t} \right)+...\right]\dot \phi
+\left[-2\pf{t'}{t}\pf{x'}{t}+2  \pf{t'}{x}\pf{x'}{x}+...
 \right]\dot\phi' \nonumber \\
&& -2il_p\left[ 3\left(\pf{x'}{t}\right)^2\frac{\pa}{\pa
    x}\left(\pf{x'}{t}\right)+... \right]\phi''-2il_p\left[
  3\left(\pf{t'}{x}\right)^2\frac{\pa}{\pa
    t'}\left(\pf{t'}{t}\right)+... \right]\ddot\phi \nonumber \\
&& -2i l_p \left[\left(\pf{t'}{t}\right)^3-\pf{t'}{t}
  \left(\pf{t'}{x}\right)^2+...
  \right]\pa_t^3\phi \nonumber \\
&&-2il_p\left[3\pf{t'}{t}\left(\pf{x'}{t}\right)^2-
2\pf{t'}{x}\pf{x'}{t}
\pf{x'}{x}-\pf{t'}{t}\left(\pf{x'}{x}\right)^2+...\right]\pa_x^2\pa_t\phi
\nonumber \\
&&-2il_p\left[
  3\left(\pf{t'}{t}\right)^2\pf{x'}{t}-\left(\pf{t'}{x}\right)^2\pf{x'}{t}
-2\pf{t'}{t} \pf{t'}{x} \pf{x'}{x}+...\right]\pa_t^2 \pa_x \phi+...\label{lp}
\eean
Covariance means that the corresponding coefficients in
\eq{fe} and  \eq{lp} be equal. For linear transformation,
$\pf{t'}{t}$, $\pf{x'}{t}$, $\pf{t'}{x}$, $\pf{x'}{x}$ are
constants. Then, the coefficients of $\phi''$, $\ddot\phi$ and
$\dot\phi'$ yield \footnote{ The skipped 
  terms in the coefficients of \eq{lp} (denoted by ``...'' ) involve
  derivatives of $\pf{t'}{t}$, 
  $\pf{x'}{t}$, $\pf{t'}{x}$, $\pf{x'}{x}$, which vanish in linear
  transformations.}
%************************clo clt cle
\bean
\left(\pf{t'}{t}\right)^2-\left(\pf{t'}{x}\right)^2 &=&1 \label{clo}\\
\left(\pf{x'}{x}\right)^2-\left(\pf{x'}{t}\right)^2 &=&1 \label{clt}\\
-2\pf{t'}{t}\pf{x'}{t}+2\pf{t'}{x}\pf{x'}{x}&=&0 \label{cle},
\eean
These three equations just give the Lorentz transformation. However,
by comparing the coefficients of $\pa_t^3\phi$,  we obtain
%******************************
\bean
\pf{t'}{t}=1,
\eean
which means that no boost is allowed for the linear
transformation.

Kimberly, {\em et al.}\cite{posi} argues that since there are plane wave
solutions, the linear contraction $p_a dx^a$ must be
invariant. Therefore,  the corresponding transformation on
position space is
%*****************************ilt ilx
\bean
dt'&=&\gamma (dt-vdx)[1+(\gamma-1)l_p E -\gamma l_p v p_x] \label{ilt}
\\
dx'&=&\gamma (dx-vdt)[1+(\gamma-1)l_p E -\gamma l_p v p_x] \label{ilx}.
\eean
Although the above relations, originally shown in \cite{posi}, are for
infinitesimal separations $dt$, $dx$, they should also hold for finite
separations. Hence, we have an energy-dependent linear transformation
on position space. However, this contradicts our conclusion that no
boost is allowed for linear transformations except for the case
$v=0$.

\section{Conclusions}
We present three distinct methods to translate the behavior in momentum
space into position space. We first identify the boost velocity with
the real velocity  measured by an inertial observer. Then we show
that any non-linear transformation in momentum space always leads to
the usual Lorentz transformation in position space. By applying
Fourier transform to a scalar field, we also obtain the same
result. However, some additional requirements  have
to be imposed. From the covariance of the modified  Klein-Gordon
equation, we show that only pure rotations are permitted among linear
transformations and the condition that $p_a dx^a$ remains an invariant
scalar is not compatible with the covariance of the field equation.

\begin{center}
{\bf  \large Acknowledgments}
\end{center}
We would like to thank Dr. S. N. Manida for helpful comment on the
original manuscript. S. Gao was supported in part by FCT
(Portugal). X.Wu thanks the support of Alexander von Humboldt
foundation and the Max-Planck Institute of gravitational physics,
Wu is also supported by National Science Council of Republic of
China under Grant No. NSC 92-2816-M-008-0004-6 .

\end{document}